# REQUIREMENTS AND THE BASELINE PLAN

**Rizwan Jameel Qureshi**
Department of Computer Science, COMSATS Institute of Information Technology, Defence Road, Lahore.
anriz@hotmail.com

*ABSTRACT: For each software project a plan is developed, according to a documented procedure, that covers the software activities and commitments. The requirements allocated to software form the basis for the software development plan. Estimates for critical computer resources are documented, reviewed, and agreed to. All affected groups and individuals understand the estimates and plans and commit to support them. Senior management reviews the estimates and plans before external commitments are made. Software risks associated with the cost, resources, schedule, and technical aspects of the project are identified and evaluated, and contingencies are documented. Planning and estimation data are collected for use in planning subsequent projects and for input in management oversight review meetings.*

## INTRODUCTION:

Project planning lays the foundation for your project management success. Quality project planning enables project managers to derive useful information for making decisions to achieve safety, time and budget goals. Poor project planning incapacitates project management from using any analytical process to achieve the same goals. It is the "garbage in, garbage out" axiom at work. Therefore, the planning stage for any project should be given as much attention as possible.

Project planning in many industries is still a "start from scratch", labor-intensive effort for every new project. Planners analyze each logical work element by defining the necessary steps (tasks), estimating their duration and resource requirements and laying out a logic network for sequencing (scheduling) them. Given sufficient time, experienced personnel can produce the highest quality plans with the manual planning process. Unfortunately, business pressures and time constraints do not always allow planners the luxury of manually planning to the best of their abilities.

With the advent of computers and project management software, re-useable project files (project templates) have become a valuable tool for the project planner. Project templates increase planner productivity and help standardize estimates from project to project. The bulk of the data entry, estimating and scheduling effort is already done. Planners can focus on customizing previous (historical) efforts.

However, the use of project templates does have some pitfalls. An organization needs to be careful to allow planners adequate time to review and customize project templates or suffer from "Cut and Paste syndrome" (CPS). CPS inevitably leads to errors such as including the wrong (old) steps, forgetting necessary new steps and overlooking schedule logic and/or man hour estimates that are not customized to the current situation.

### Project Planning with Estimating Modules

Dynamic estimating modules incorporate the benefits of project templates while solving their potential pitfalls. Dynamic estimating modules can be thought of as large, flexible project templates that incorporate many different scenarios. Estimates and resource requirements are stored as estimating formulas instead of fixed amounts. Schedule logic is "super-networked" to allow flexibility in including or excluding ranges of tasks while maintaining the logic network integrity.

To use dynamic estimating modules, planners enter the required parameters into the project estimating software. The appropriate task lists, durations, resource assignments and predecessor schedule logic are extracted (and calculated) from the estimating modules in accordance with the supplied parameters and exported into the main project file.

### Evolution of the Project Planning Process

Over time the benefit of using dynamic estimating modules will multiply. Since estimating formulas and schedule logic are captured within the estimating modules, there will be no drop-off in quality when new or inexperienced planners arrive on the job. Dynamic estimating modules will allow larger companies the means to standardize planning and estimating using their best practices across the enterprise. Dynamic estimating modules are the next evolution in the refinement of the project planning process.

The Project Team Kick-off Meeting is the first meeting with the project team members to discuss the project and the work that will be completed. Prior to the meeting, the Project Definition Document or Project Proposal should be distributed to the team members. This meeting introduces the members of the project team and provides the opportunity to discuss the role of each team member in the project work. The other base elements in the project may also be discussed at this meeting (Project Task List, Schedule, Communication Plan, etc.).

**Purpose**
- Introduce the project team members.
- Discuss the general project and address any questions.
- Discuss the base project plan elements: Project Task List, Schedule, Communication Plan, meeting schedules, project administration, etc.

**Benefits**
- All project team members are aware of the complete project.
- Defines the parameters of the project work, timeline, and control process.
- Identifies any missing team members or technical stakeholders.
- Meeting Guidelines
- Pass out an agenda at least one day in advance.
- Invite project participants who are not full time team members (from support organizations), if appropriate, schedule items pertinent to them early in the meeting so they can leave after these items are covered if they wish.

- Begin and end on time.
- Ask someone to volunteer to take meeting notes and distribute to the project team following the meeting.
- If a topic comes up that is too big to handle in the time frame allotted, schedule another time to deal with it.
- Draw people out. Do not assume silence is consent.
- Create an action list with responsibilities assigned and check the list for completion at the next meeting.

**Communication Plan**

A communication plan identifies people with an interest in the project (stakeholders), communication needs, and methods of communication. Communication planning helps to ensure that everyone who needs to be informed about project activities and results gets the needed information.

- Who develops a Communication Plan?
- The project manager is responsible to identify communication needs and to decide whether a formal communication plan is needed.

**What kinds of projects need a Communication Plan?**

Although every project undergoes some kind of communication planning, it is frequently informal - determining who needs to attend which meetings, receive which reports, etc. Projects of long duration will benefit from formal planning because the project stakeholders are likely to change over time. Projects that affect a large number of people or organizations may also benefit from formal planning to ensure full identification of both stakeholders and of communication needs.

**Guidelines**

A communication plan needs to consider, and where appropriate document each of the following items:

- Project Name
- List of Stakeholders (Who has interest in the project? See the project definition for an initial list of stakeholders. Be sure to include both business and technical stakeholders.)
- Information Needs (What kinds of information about the project are of interest? Consider need to communicate plans, status and progress reports, changes, major events, availability of prototypes and demonstrations, etc.)
- Communication Methods (What information will be communicated to what groups in what ways? Common methods include reporting and documentation, email, meetings, and web sites.)

A software life cycle with defined stages is used as a framework for the plan. The plan includes documented size, cost, schedule, and resource estimates developed according to a written procedure and based on historical data. The main objective of this paper is to realize the developers about the importance of baseline plan before they made a detailed analysis to develop a system. It is important to develop a baseline plan for the developer by providing a boundary to approve or reject the project. The detailed analysis is only started if the customer and software house mutually approves baseline plan [2]. How it works? The main purposes of developing the baseline document are:
- working of current system;
- problems of the current system;
- the scope of new system;
- goals of new system,
- benefits achieved to customer after implementing the new system;
- budget required to develop the new system;
- duration required to develop the new system;
- budget required to implement the cost;
- handle the management issues.

Customer can compare costs required to develop and implement the system vs. the benefits will achieve after implementing the new system. He will verify the functions and duration.

Software houses can use baseline document as a guide to find detailed relevant requirements during the analysis phase. There is no screening process available to developer to check that the requirements gathered are relevant or vague. Also, there is no surety that gathered requirements are complete. It also helps the analyst to face unexpected responses of the investors after release of the software systems; such as, this system is not functioning up to expectations, this is not completing all requirements, system development is exceeding the budget and duration and it is consuming much more resources as expected.

**Main Communication Questions**

The customer must answer three main questions during the communication.

- What are the main business needs or deeds?
- Who are going to use the application?
- What are the main functionalities of the new system?

**Importance of Baseline plan**

There are so many reports that customers asked the developers to develop a system that increase their business and also mentioned the duration for the completion of the projects. The development team asked the customer about the main benefits and goals of the system. Customers were unable to explain about the main benefits and goals of the systems and provided some vague and ambiguous requirements. The development teams have to spend extra efforts to extract the requirements and completed the projects within the deadlines. Customers normally made complaints after installment that these systems are not improving their businesses and failing to meet their expectations and put more requirements to add in to the systems.

Baseline plan plays an important role for the success or failure of a project. Baseline plan is like the foundation of a building. The building will collapse after few years if the foundations are weak. SW will be deteriorated after few years if the baseline plan is not properly worked and documented.

**Signs of a Quality Baseline Plan**

These are the signs of a quality baseline plan.
- Developers did understand the complete requirements of the new system.
- Project scope is clearly defined.
- Working of the current system is completely understood.
- Project team has these characteristics.
  - Competent
  - Self organized
  - Problem solving
  - Decision making
  - Common scope

- o Collaboration
- o Trust and Respect among the team members
- o Flexible
- Platform independent tool must be selected to develop the system.
- Project schedule must be carefully evaluated.
- Latest and widely accepted process model should be used.
- Customer must be involved throughout the preparation of a baseline plan and throughout the system development.
- Consistent work is required to prepare a quality baseline plan.
- Keep it simple (KIS) principle must be followed to prepare a baseline plan and for the whole development.
- Risks must be carefully evaluated.
- Develop use-case diagrams initially.
- Rough architecture of the proposed system must be developed.
- Whole SW must be planned to deliver within couple of months.
- SW must be planned to develop in increments.
- Request the customer to provide an index value against each increment.
- Feasibility assessments for each increment are prepared.
- Risk assessments for each increment.
- Customer is requested to split the requirements if an increment is consuming more than three weeks.
- Feasibility assessments are again made.
- Risk assessments are again made.
- Requirements with high index values are planned to develop first.
- Requirements with high-risk values are planned to develop first.
- Factor of Reuse must be in mind during the planning.
- Define user categories.

A baseline plan is a systematic and pragmatic approach to direct and complete the project. It provides a framework to make a project profitable in terms of development effort and investment [1]. Baseline plan is one of the project drivers. This type of issues cannot be resolved without defining a baseline plan; such as, main requirements to achieve desired benefits, number of people, source line of code, effort, budget and schedule?

Project planning begins with the problem definition phase of the project. Problem definition means finding the differences between the current situation and desired or expected situation of the SW system. A senior manager or business group or information system manager or a committee identifies the main problems of organization such as COMSATS Institute of Information Technology Lahore campus needs SW those can be used to schedule the timetable, attendance, result generation and registration. These all are different projects. The SW projects are classified or ranked based on the project size, cost, schedule, resources, technologically, ease of implementations and how much it will add value to the organization. A system service request (SSR) is made to analyst or project director after selection of the most desired project having low cost, less time and easiest to implement by the senior manager or business group or information system manager or a committee during the project initiation phase.

The analyst is a person who is responsible to analyze, design and develop the SW systems. System analyst will submit the baseline plan. The main activities of the planning phase are as follows.
- Review the baseline plan.
- Approve or revise or refuse the baseline plan.
- Priorities the baseline plans for different projects.
- Resources are allocated, such as money, people and equipments for the approved project.
- A project team is developed for the approved project.

**How to define a baseline plan?**
Baseline plan is composed of these main contents:
- project Introduction;
- feasibility assessments;
- management issues.

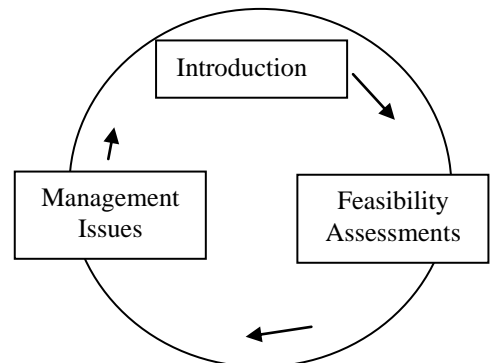

Fig 1 Diagrammatic View of IFM Model

**Project introduction** is composed of working of the current system, problems of current system, title of new system, scope and goals of new system. **Scope** helps the developers to find the main requirements. **Goals** are the main functionalities of the system. **Feasibility assessments** help the analyst to estimate the total benefits and costs in terms of costs and benefits analysis sheet. There are four important feasibility assessments are made.
- Economical
- Technical
- Operational
- Schedule

**Economical feasibility** helps to measure the project benefits and costs. Project benefits are measured in terms of tangible and intangible. Tangible benefits are the benefits that are measurable in terms of monetary values; such as, time saving, cost saving, error reduction, opening new market or sales opportunities and better management planning and control. Intangible benefits are the benefits that cannot measure in terms of monetary values; such as quality, efficiency, employee morale and customer goodwill. Analyst also mentioned one time and recurring costs. One time cost displays money used to develop the system. Recurring costs display money used to evolve the software and expand the system. Software is evolved due to new business needs or deeds, new customer requirements, business growth and down sizing.

Maintenance is an unwanted variable involved when Software is evolved. Corrective, perfective, adaptive and preventive are four main types of maintenance. Corrective maintenance is done so that system can work correctly on the basis of defects reported by the customer after usage. Perfective maintenance is done to complete the system in terms of missing functionalities. Adaptive maintenance is done so that Software can run on a new Operating System other than that was used to develop the system. For example, Java was used to develop an online registration system. Customer was using Microsoft Windows Operating system when software was installed. Customer decided to switch the Operating System from Microsoft Windows to Linux after installation of the software. Maintenance is done so that application made in Java can adapt new operating system. Preventive maintenance is done to improve the quality of the software. Quality of the Software is reduced as preventive, perfective and adaptive maintenances are made. Project cost estimation methods are used to measure the total project cost using Function Point, Constructive Cost Models and Backfiring methods. These cost estimation methods help the developers to find the development effort, number of people, source line of code and duration of the project.

**Technical feasibility** helps to measure the hardware and software cost needs to develop the system. Technical feasibility also includes the technological cost in terms of learning the tool and platform. **Operational feasibility** is used to measure the operational benefits and operational cost needs to implement the system. **Schedule feasibility** is used to show the total duration and breakdown of the tasks.

**Management issues** help to handle the **stakeholders, market trends, support** of the top management and **risks management** and **configuration management**. Stakeholders are the sources of requirements. Project must meet the market trends at the completion. Project must not loose the support of top management during completion.

**Criteria To Approve the Baseline Plan For Discretionary Projects**
- How much the new system will save costs?
- When and where cost will be saved?
- How much new system will improve the organizational processes and information processing?
- How much better services are provided to customers?
- Will the system implement within affordable time?
- Is the organization having required resources?
- The benefits are long term or short term.
- Long-term benefits are covering the development cost or not.

Discretionary and non-discretionary are two main types of projects. The criteria mentioned above implements on discretionary projects.

**Main benefits of the Baseline Plan**
The answers of the above mentioned questions would help the analyst to prepare a sound baseline plan. The main benefits of the baseline plan are as follows.

- To uncover main functional requirements.
- Track the project in the correct direction.
- To remove all misunderstandings and uncertainties regarding the project. Stakeholders.
- Users can makeup their mind before a detailed analysis is made.
- To maintain a feedback loop among the customer, analyst and the developer.
- It is easy for the analyst to trace and manage the effects on the base line plan whenever a change is made in the project scope or goals [3].

## CONCLUSION

A base line plan provides a channel for the analyst and investor to approve or refuse the project on the basis of cost benefit analysis sheet. Baseline plan is also helpful for the developer by estimating the all-possible risks. SW project management depends a lot on an effective baseline plan. Baseline plan may change as the project development progresses, such as during the analysis phase after gathering the detailed requirements. Analyst must measure all possible problems that may arise during the system development. He should also prepare the strategies to solve those problems. Baseline plan is a driver for driving the project. Baseline plan is considered iterative and it is completed only when the project is completed. Baseline plan is revised during the different phases of system development because more detailed information about the project is available as the project progresses. It is compulsory to change the baseline plan throughout the phases of the system development to make the project successful.

**References:**

1. Suzzane Robertson (Principal and founder of Atlantic System Guild), "Requirements and the Business Case", IEEE Software, Sep/Oct. 2004, pp. 93-95.
2. (Jeffrey A. Hoffer 1999) Jeffrey A. Hoffer, Joey F. George, Joseph S. Valacich. "Modern Systems Analysis & Design". Addison-Wesley, USA, 1999.
3. (Roger S. Pressman 2005) Roger S. Pressman. "Software Engineering", McGraw Hill, USA, 2005.